\documentclass[prb,aps,twocolumn,nofootinbib,groupaddress]{revtex4-1}
\usepackage{latexsym}
\usepackage{graphicx}
\usepackage{verbatim}
\usepackage{multirow}
\usepackage{amsmath}
\renewcommand{\vec}[1]{{\mathbf{#1}}}
\newcommand{\beq}{\begin{eqnarray}}

\newcommand{\eeq}{\end{eqnarray}}
\usepackage{mathrsfs}
\usepackage{float}
\usepackage[usenames, dvipsnames]{color}
\usepackage{mathtools}
\usepackage{slashed}
\usepackage{physics}	
\usepackage{graphicx}   
\usepackage{subfigure}
\usepackage[hidelinks]{hyperref}   
\usepackage{bbold}
\usepackage{wasysym}

\usepackage{footnote}

\begin{document}

\title{Extracting correlation effects from Momentum-Resolved Electron Energy Loss Spectroscopy (M-EELS): Synergistic origin of the dispersion kink in Bi$_{2.1}$Sr$_{1.9}$CaCu$_2$O$_{8+x}$}
\author{Edwin W. Huang}
\affiliation{Department of Physics and Institute for Condensed Matter Theory, University of Illinois at Urbana-Champaign, Urbana, Illinois, 61801}
\author{Kridsanaphong Limtragool}
\affiliation{Department of Physics, Faculty of Science, Mahasarakham University, Khamriang Sub-District, Kantharawichai District, Maha-Sarakham 44150, Thailand}
\author{Chandan Setty}
\affiliation{Department of Physics, Rice University, Houston, Tx. 77005}
\author{Ali A. Husain}
\affiliation{Department of Physics and Astronomy and Quantum Matter Institute, University of British Columbia, Vancouver, BC V6T 1Z4, Canada}
\author{Matteo Mitrano}
\affiliation{Department of Physics, Harvard University, Cambridge, Ma. 02138}
\author{Peter Abbamonte}
\affiliation{Department of Physics, University of Illinois at Urbana-Champaign, Urbana, Illinois, USA}
\author{Philip W. Phillips}
\affiliation{Department of Physics, University of Illinois at Urbana-Champaign, Urbana, Illinois, USA}

\begin{abstract}
We employ Momentum-Resolved Electron Energy Loss Spectroscopy (M-EELS) on Bi$_{2.1}$Sr$_{1.9}$CaCu$_2$O$_{8+x}$ to resolve the issue of the kink feature in the electron dispersion widely observed in the cuprates. To this end, we utilize the GW approximation to relate the density response function measured in in M-EELS to the self-energy, isolating contributions from phonons, electrons, and the momentum dependence of the effective interaction to the decay rates. The phononic contributions, present in the M-EELS spectra due to electron-phonon coupling, lead to kink features in the corresponding single-particle spectra at energies between 40 meV and 80 meV, independent of the doping level. We find that a repulsive interaction constant in momentum space is able to yield the kink attributed to phonons in ARPES. Hence, our analysis of the M-EELS spectra points to local repulsive interactions as a factor that enhances the spectroscopic signatures of electron-phonon coupling in cuprates. We conclude that the strength of the kink feature in cuprates is determined by the \emph{combined} action of electron-phonon coupling and electron-electron interactions.\end{abstract}

\maketitle
\section{Introduction}
In strongly correlated electron matter, the non-interacting band dispersion fails to describe the elementary excitations.  The departure from non-interacting physics is usually captured by the self-energy.  It is in this context that the kink-like feature measured from angle-resolved photoemission spectroscopy (ARPES)~\cite{Damascelli2003, Kaminski2000, Puchkov1996, Hwang2007, Hwang2004, Lanzara2001} at 60meV has risen to the fore as a tell-tale signature of a possibly universal energy scale in cuprate physics over a wide range of doping.   Although phonons\cite{Lanzara2001,zaanenprl} are widely cited as the origin of the kink, there is good reason to believe that phonons alone are insufficient\cite{Giustino2007,Chubukov2004,Kordyuk2006,Dahm2009,Raas2009,Matsuyama2017}.  To help resolve this puzzle, we resort to  Momentum-Resolved Electron Energy Loss Spectroscopy (M-EELS) which provides a measurement of the 2-particle response or the density-density response function. Applying standard many-body approaches to the M-EELS data allows us to disentangle the many-body excitations encoded in the electron self-energy.


An advantage M-EELS has over ARPES is that  because ARPES measures occupied states, extracting the self-energy from ARPES data relies on additional assumptions concerning the unoccupied spectral function, which may or may not be justified. A complementary approach to determining the self-energy from experimental data is to consider the scattering of electrons from bosonic fluctuations. Formally, two-particle response functions that characterize bosonic fluctuations can be related exactly to the single-particle self-energy via Hedin's equations\cite{Hedin}. This relationship allows one to identify features in the single-particle spectra with particular aspects of the two-particle spectra.  
To do so, we employ the GW approximation which is based on assuming a bare vertex function in Hedin's equations. The self-energy is given in imaginary time by
\begin{equation}
\Sigma(\vec{k},\tau) = \int \frac{\dd{\vec{q}}}{(2 \pi)^2} G(\vec{k}-\vec{q},\tau) W(\vec{q},\tau + 0^+) \label{eq:GWtau}
\end{equation}
where $\tau$ is imaginary time, $G$ is the Green function and
\beq\label{eq:W}
W(\vec{q},\tau) &= V(\vec{q})\delta(\tau) + V(\vec{q}) \chi(\vec{q},\tau) V(\vec{q})
\eeq
is the screened Coulomb interaction, in terms of the bare Coulomb interaction $V(\vec{q})$ and the charge susceptibility $\chi(\vec{q},\tau)$. In our calculations, the susceptibility will be taken directly from momentum resolved electron energy loss spectroscopy (M-EELS) measurements of the density-density response function in the high-$T_c$ cuprate Bi$_{2.1}$Sr$_{1.9}$CaCu$_2$O$_{8+x}$ (BSCCO). This recently developed technique~\cite{Egerton2007} provides reliable measurements of the total density response of a system for momenta throughout the Brillouin zone with meV resolution.  Moreover, unlike other probes (like Inelastic X-ray Scattering or IXS), the density response of M-EELS primarily originates from valence electrons and shields out contributions from the core states.  Equipped with experimental knowledge of the density response of valence electrons for all frequencies and momenta, we undertake the tasks of determining the corresponding self-energy and examining to what extent the results agree with other known probes of the self-energy.


In this paper, we showcase how M-EELS can be used as a purely non-optical probe of correlation effects captured in the momentum and frequency dependent scattering rate or self energy. Using the density response measured in M-EELS~\cite{Vig2017,Mitrano2018,Husain2019}, we employ the GW method to evaluate the self-energy for the under-, optimally-, and over-doped copper oxide superconductor Bi$_{2.1}$Sr$_{1.9}$CaCu$_2$O$_{8+x}$ (BSCCO). In the process, we isolate the contributions from phonons as a cause of the kink in the energy dispersion in the momentum and energy-dependent curves as seen in ARPES.  Hence, we are able to offer new insights into the origin of the debated kink features around $\sim60meV$~\cite{Damascelli2003, Kaminski2000, Puchkov1996, Hwang2007, Hwang2004, Lanzara2001}. Our calculations find that, independent of doping, kinks appear at the energies of the phonons visible by M-EELS when considering effective electron interactions that are local in real space.




\section{Preliminaries} 
\begin{figure}
	\centering
	\includegraphics{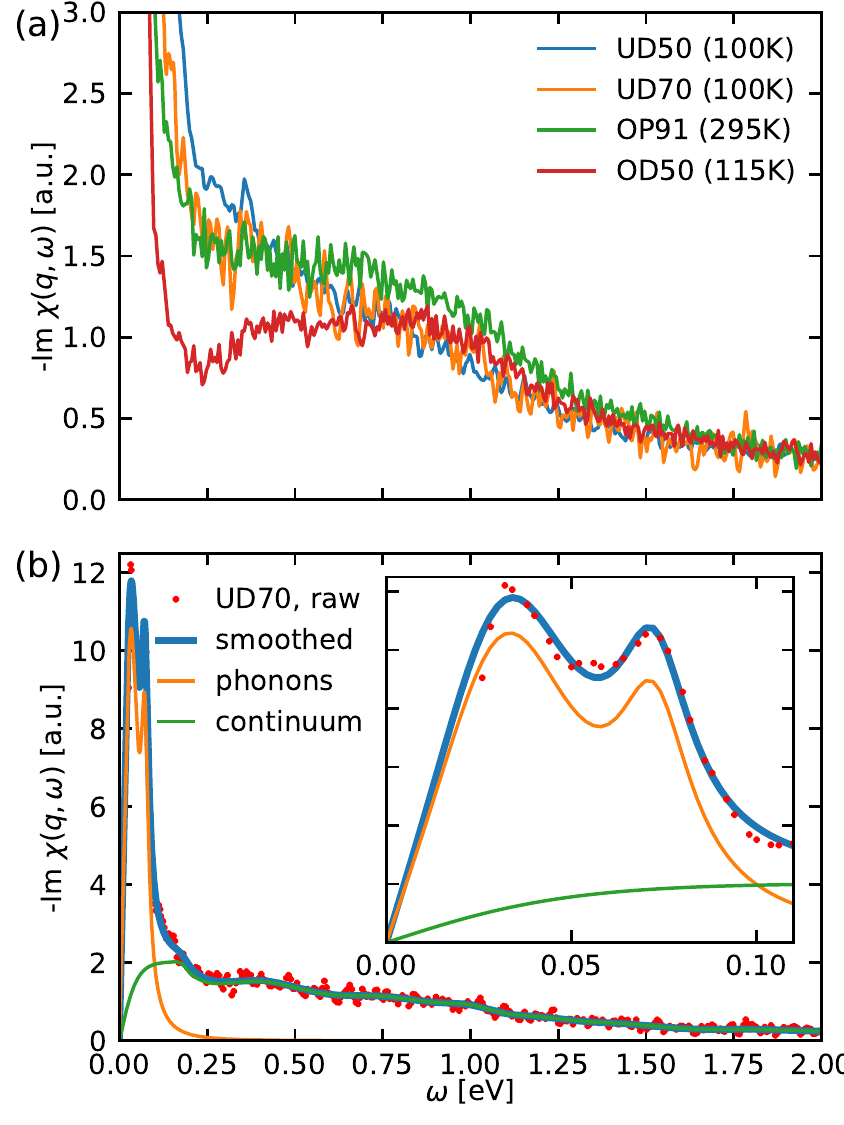}
	\caption{(a) Plots of the imaginary part of the density-density response function at momentum $q = (0.5,0)$ in r.l.u., as measured by M-EELS for four dopings of BSCCO. (b) Separation of the M-EELS data for UD70 into phonon peaks (orange) and the electronic continuum (green), which add to form the smoothed curve (blue). The phonon peaks are fitted with two anti-symmetrized Lorentzians, although they may consist of more than two phonons that are not resolved. The smoothed curve is obtained by a smoothed spline fit to the raw data (red dots). Inset: same plot at low frequencies.}
	\label{fig:fig2}
\end{figure}

The imaginary part of the density response function, $\chi''(\vec{q},\omega)$, in BSCCO can be measured from M-EELS as previously reported in \cite{Vig2017,Mitrano2018,Husain2019}. In this Letter, we use $\chi''(\vec{q},\omega)$ measured in BSCCO at four different doping concentrations to compute the imaginary part of the self-energy. The four dopings include underdoping with $T_c = 50$ K and $T_c = 70$ K, optimal doping with $T_c = 91$ K, and overdoping with $T_c = 50$ K. The generic features of $\chi''(\vec{q},\omega)$ as shown in Fig.~\ref{fig:fig2}b include two phonon peaks at energies about 40 meV and 70 meV and a broader electronic continuum with an edge at about 1 eV. The plots of $\chi''(\vec{q},\omega)$ for the four dopings at various values of $q = |\vec{q}|$ as described previously\cite{Husain2019}. In Fig.~\ref{fig:fig2}, we display $\chi''(\vec{q},\omega)$ for all four dopings at fixed $q = 0.5$ reciprocal lattice units (r.l.u.) to demonstrate the doping-dependent feature of $\chi''$. Comparing to the flat background from 100 meV to 1 eV in the optimally doped case, there is a suppression (an enhancement) of $\chi''$ at about 200 - 400 meV in the overdoped (underdoped) case.

We assume that $\chi''(\vec{q},\omega)$ does not depend on the direction of $\vec{q}$, i.e., $\chi''(\vec{q},\omega) = \chi''(q,\omega)$. Ref. \cite{Mitrano2018} showed that $\chi''(\vec{q},\omega)$ measured along the nodal and anti-nodal directions coincide for $q \equiv |\vec{q}|$ between 0.1 to 0.5 r.l.u. For smaller values of $q \approx 0.05$ r.l.u., there is a difference between $\chi''$ in the two directions at energies below 1 eV. For simplicity of the calculations, we ignore this deviation at small $q$ and take $\chi''(\vec{q},\omega) = \chi''(q,\omega)$ for all $\vec{q}$ in the Brillouin zone.

To obtain the screened interaction $W$ from the density response $\chi$, a concrete form for the electron interaction, $V(q)$, is required.  Ref. \onlinecite{Mitrano2018} considered the susceptibility $\chi(q,\omega)$ in terms of the background susceptibility $\varepsilon_\infty$ (4.5 for BSCCO \cite{Levallois2016}) and polarizability $\Pi(q,\omega)$
\beq
\chi(q,\omega) = \frac{\Pi(q,\omega)}{\varepsilon_\infty - V(q)\Pi(q,\omega)},
\eeq
and found that the imaginary part of the polarizability factors in momentum and energy with the form (for optimal doping)
\beq
\Pi''(q,\omega) = -\Pi_0(q)\tanh\frac{\omega_c^2(q)}{\omega^2}, \label{eq:Pifactor}
\eeq
if the effective interaction is given by
\beq
V(q) = V_0\frac{\exp(-z q)}{q}, \label{eq:Vq}
\eeq
where $V_0 = 820$ $eV$ $\mathring{A}^3$ and $z = 14.3$ $\mathring{A}$. The prefactor function $\Pi_0(q) \propto q^2$ and the cut-off frequency $\omega_c(q) \approx$  1 eV. The specific form of $V(q)$ here is not determined directly or even physically motivated, but chosen based on a fitting analysis that assumes factorizability of the polarizability. Notably, even by this scheme, its value for large $q$ is not strongly constrained. We therefore consider also $V(q)$ with a more regular momentum dependence. Specifically, we will study the simplest case in which $V(q) \propto 1$, corresponding to a local real-space interaction.  One of our main conclusions is that while $V(q)$ as given in \eqref{eq:Vq} leads to factorizability of $\Pi(\vec{q},\omega)$, it does not produce the expected behavior of the kink in the spectral function $A(\vec{k}, \omega)$ in our calculations.

For the band structure, we include up to next-next nearest neighbor hoppings, $t$, $t'$, and $t''$:
\begin{align}\label{eq:dispersion}
\epsilon_{\vec{k}} = & -2t(\cos(k_x a)+\cos(k_y a)) - 4t'\cos(k_x a)\cos(k_y a) \nonumber \\
&- 2t''(\cos(2 k_x a) + \cos(2 k_y a)).
\end{align}
The hoppings we use in this paper are $t = 0.42 eV$, $t' = -0.110 eV$, and $t'' = 0.055 eV$ \cite{Pavarini2001,Markiewicz2005}. The lattice parameter is $a = 3.81$ $\mathring{A}$  [\onlinecite{Mitrano2018}].

One of our goals is to separate and isolate the contributions from phonons and from the electronic background to the self-energy. To this end, we fit the two phonon peaks at energies about 40 meV and 70 meV with two Lorentzian functions (antisymmetrized to obey $\chi''(\vec{q},-\omega) = -\chi''(\vec{q},\omega)$). These peaks may be subtracted from the data to obtain the electronic continuum, as shown in Fig.~\ref{fig:fig2}b. We then compute the imaginary part of the self-energy from the density response considering only the phonons peaks, only the electronic continuum, or the entire spectrum. Plugging in the self-energy to Dyson's equation yields the Green function, from which we plot the spectral function and visualize the dispersion by looking at the maxima of momentum distribution curves (MDCs), as is commonly done to analyze ARPES data.

As is evident from Eq. (\ref{eq:W}), there are two contributions to the self energy. The first term that contains only the bare interaction is frequency-independent and hence just provides a shift to the band dispersion. This term will be dropped in our calculations to avoid double-counting, since our non-interacting dispersion \eqref{eq:dispersion} was determined by a fit to experimental ARPES data, which of course include the effects of screening. The frequency dependence of the self-energy arises entirely from the second term that includes the density-density response. Then, in real frequency, \eqref{eq:GWtau} takes on the form
\begin{align} \label{eq:Sigma_gamma0}
\Sigma''(\vec{k},\omega) &= \int \frac{\dd{\vec{q}}}{(2 \pi)^2} \dd{\Omega} V(\vec{q})^2 \qty[f(-\Omega) + n(\omega - \Omega)] \nonumber\\
&\times\frac{-1}{\pi} G''(\vec{k}-\vec{q}, \Omega) \chi''(\vec{q}, \omega - \Omega) 
\end{align}
where $f(\omega)$ and $n(\omega)$ are the Fermi and Bose distribution functions, respectively, and $\Sigma$, $G$, and $\chi$ are understood to be evaluated with an infinitesimal displacement above the real frequency axis. For numerical stability, we perform all calculations with a small finite amount $\gamma$ above the real frequency axis rather than using Eq.~\ref{eq:Sigma_gamma0}. Following the methods of [\onlinecite{Schmalian1996}], $\Sigma(\vec{k}, \omega + i \gamma)$ can be evaluated efficiently via fast Fourier transforms in terms of $G$ and $\chi$ using
\beq \label{eq:Sigmart}
\Sigma(\vec{r},\omega+i\gamma) &=& \int_0^\infty \dd{t} \Sigma(\vec{r},t) e^{i(\omega+i\gamma)t}\nonumber \\
\Sigma(\vec{r},t) &=& i 2 \pi T \Re \widetilde{\chi}(\vec{r},0 + i0^+)e^{\gamma t} \rho(\vec{r},t)\nonumber \\
&-& i(2\pi)^2\nu(\vec{r},t)\left(\mathcal{A}(\vec{r},t) + \mathcal{A}(\vec{r},-t)^* e^{2\gamma t}\right)\nonumber\\
&+&i(2\pi)^2\rho(\vec{r},t) \left(\mathcal{B}(\vec{r},t)^* + \mathcal{B}(\vec{r},-t)\right)e^{2\gamma t}\nonumber\\
\eeq
with
\begin{align} \label{eq:Sigmart_defs}
\rho(\vec{r},t) &= \int_{-\infty}^\infty \frac{\dd{\omega}}{2\pi} e^{-i\omega t} \frac{-1}{\pi} G''(\vec{r},\omega+i\gamma)\nonumber \\
\mathcal{A}(\vec{r},t) &= \frac{i}{2\pi} \int_{-\infty}^\infty \frac{\dd{\omega}}{2\pi} e^{-i\omega t} f(\omega+i\gamma)^* G(\vec{r},\omega+i\gamma)^*\nonumber \\
\nu(\vec{r},t) &= \int_{-\infty}^\infty \frac{\dd{\omega}}{2\pi} e^{-i\omega t} \frac{-1}{\pi} \widetilde{\chi}''(\vec{r},\omega+i\gamma)\nonumber \\
\mathcal{B}(\vec{r},t) &= \frac{i}{2\pi} \int_{-\infty}^\infty \frac{\dd{\omega}}{2\pi} e^{-i\omega t} n(\omega+i\gamma)^* \widetilde{\chi}(\vec{r},-\omega+i\gamma) \nonumber \\
\widetilde{\chi}(\vec{r},\omega) &= \frac{1}{N} \sum_{\vec{q}} e^{i \vec{q} \cdot \vec{r}} V(\vec{q})^2 \chi(\vec{q},\omega).
\end{align}
All momentum integrals are discretized with on a grid of size $N=200\times200$. Frequency integrals are discretized with steps of $0.002 eV$ in the range $[-12eV,12eV]$. We have checked that increasing the density of the momentum and frequency grids and extending the range of the frequency integration do not affect our results. In the Fermi and Bose distribution functions, we have assumed a temperature of $T = 0.002eV$. As this value is smaller than the resolution of the M-EELS spectra, varying it does not affect our results. Finally, the parameter $\gamma$ must satisfy $\gamma < \pi T/2$ as discussed in [\onlinecite{Schmalian1996}]. We found $\gamma = \pi T/3$ to yield numerically stable results.

The calculation of the self-energy begins by using a non-interacting Green function $G(\vec{k}, \omega + i \gamma) = [\omega + i \gamma - \epsilon_{\vec{k}}]^{-1}$. After a Fourier transform to real space, Eqs.~\ref{eq:Sigmart} and \ref{eq:Sigmart_defs} are used to evaluate $\Sigma(\vec{r}, \omega + i \gamma)$, which is then transformed back into momentum space. A new Green function is obtained via
\begin{equation}
G(\vec{k}, \omega + i \gamma) = \frac{1}{\omega + i \gamma - \epsilon_{\vec{k}} - \Sigma(\vec{k}, \omega + i \gamma)}
\end{equation}
and may be used as the new input to the calculation. This procedure may be stopped after the first calculation, which we will refer to as the one-loop calculation (similar to Ref.~\onlinecite{Vig2017}), or repeated until the self-energy and Green function converge i.e. self-consistently. Typically, self-consistency is achieved within $\sim 10$ iterations.

\section{Results}

\begin{figure}
	\centering
	\includegraphics[scale=0.85]{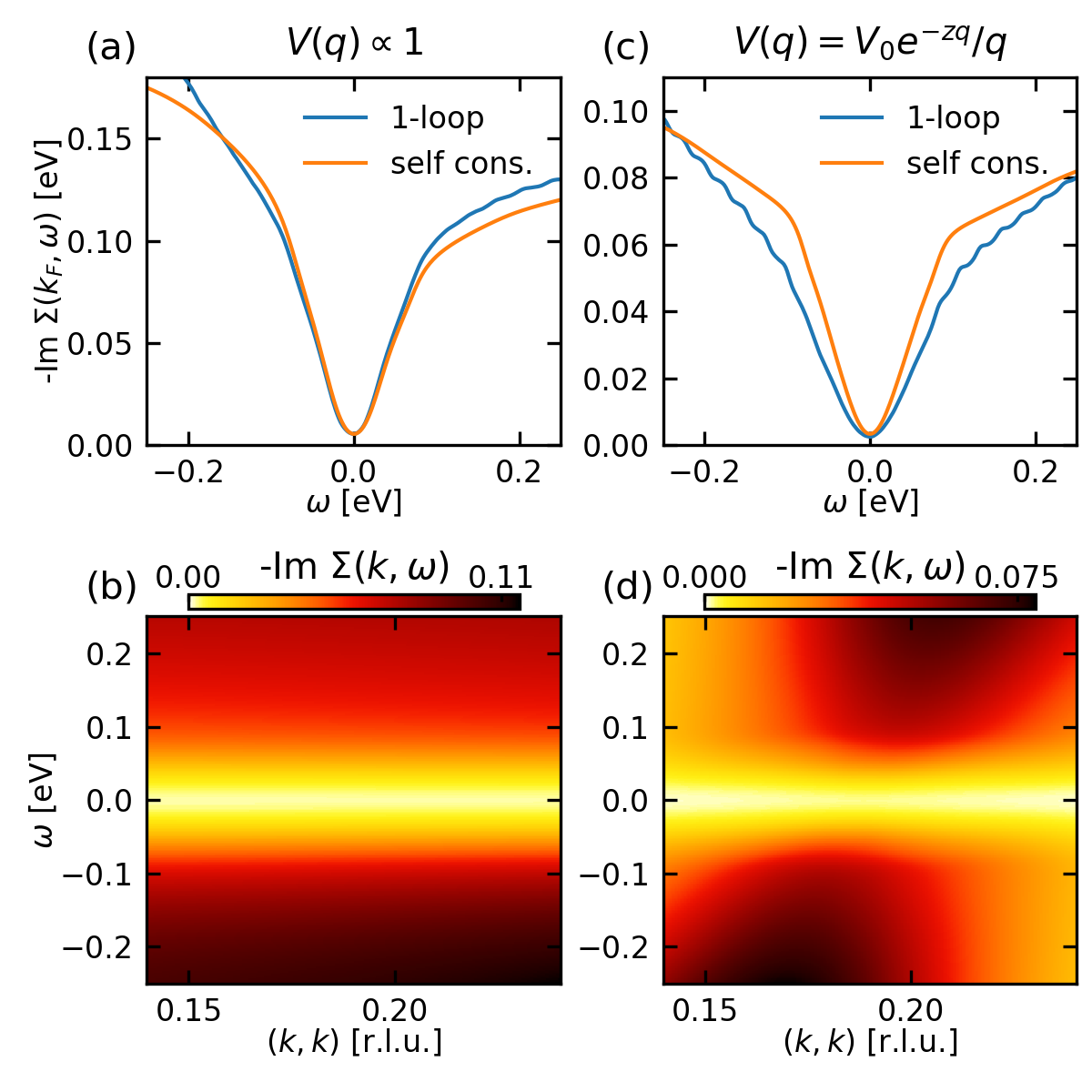}
	\caption{Plots of the imaginary part of the self-energy in the nodal direction as computed by the GW approximation with M-EELS data for underdoped BSCCO ($T_c = 70$ K).  Figs (a) and (b) were computed using $V(q)\propto 1$ whereas for (c) and (d) Eq. \ref{eq:Vq} was used.  Both (b) and (d) are 2d plots of the self-consistent calculation in which the color shows the magnitude of the imaginary part of the self-energy along a nodal cut. Darker colors correspond to greater intensity.}
	\label{fig3}
\end{figure}

We numerically calculated the real and imaginary parts of the self-energy using Eq. \ref{eq:Sigmart} and Eq. (\ref{eq:Sigmart_defs}). Figs. (\ref{fig3}a) and (\ref{fig3}c) compare the self-consistent and 1-loop evaluations of the imaginary part of the self energy for two kinds of interaction potentials, a constant $V(q)\approx 1$ and the form $V(q)=V_0 e^{-zq}/q$ as discussed in Ref.~\cite{Mitrano2018,Husain2019}. The 1-loop self-energies show minor oscillations due to the finite $200\times200$ momentum grid; these artifacts go away with increasing momentum resolution and hence can be ignored.  While the self-consistent solution and the one-loop calculation exhibit somewhat different slopes at high energy, the behaviour at low-energies is qualitatively similar. In the subsequent calculations, we do not find major differences between the two schemes, and we will focus on data from the self-consistent calculations.

We see also that the two different form of potentials $V(q)$ yield quite similar imaginary parts of the self energies, which is surprising given that one is constant and the other is sharply peaked as functions of $q$. However, as we will see, this agreement is misleading as significant changes will become apparent in our evaluation of the spectral function and dispersion curves.  Figs. (\ref{fig3}b) and (\ref{fig3}d) contain a 2d color plot of $\Sigma''(\vec{k},\omega)$ as a function of both frequency and momentum, as obtained from the self-consistent calculation.  As expected, the momentum dependence of the self-energy is stronger for the momentum-dependent interaction potential. What our analysis shows thus far is that we have a numerically self-consistent stable method to calculate the self-energy and analyse the results for an arbitrary potential that goes beyond the 1-loop approximation used previously\cite{Vig2017}.

\begin{figure}
	\centering
	\includegraphics[scale=0.85]{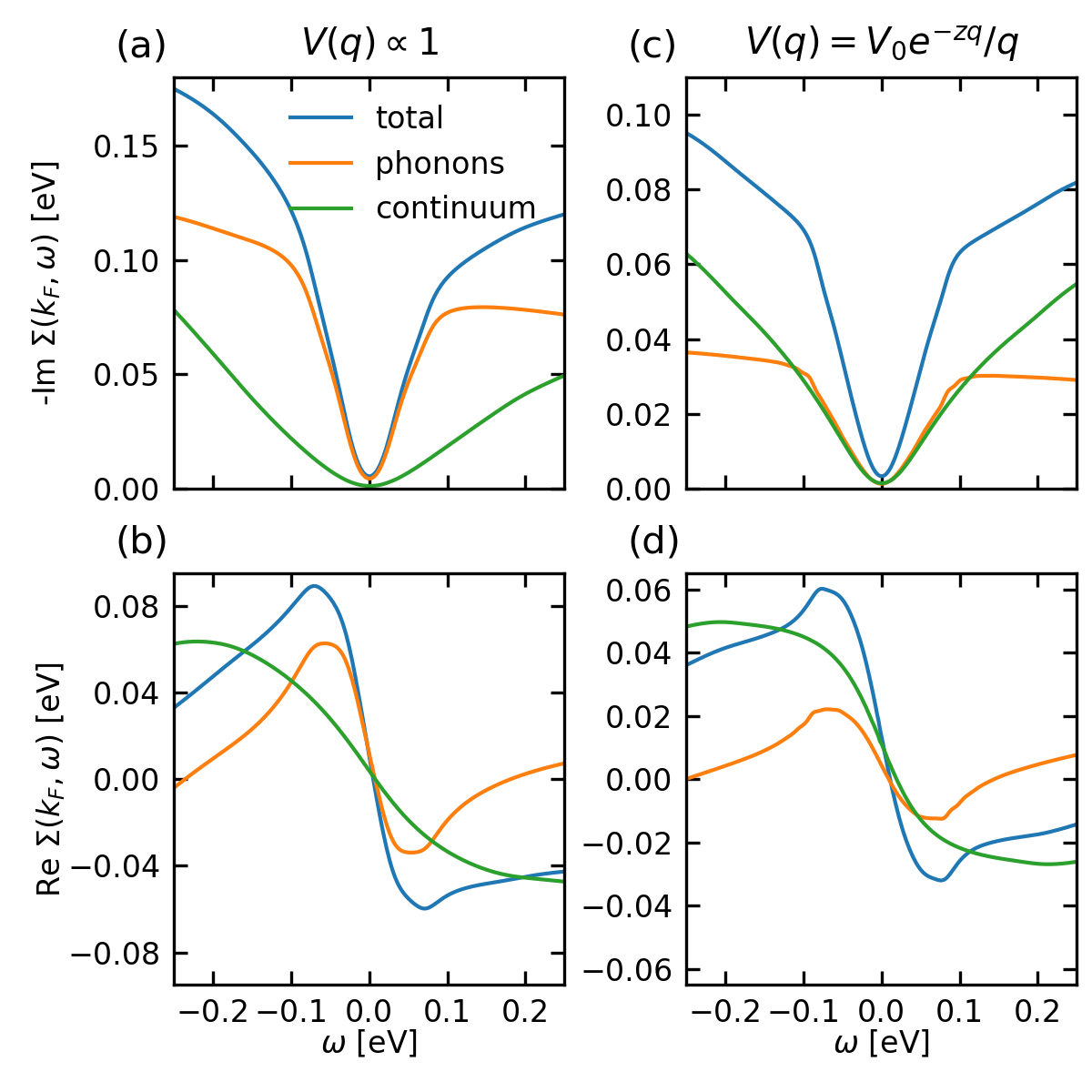}
	\caption{Imaginary (a) and (c) and real (b) and (d) parts of the self-energy for UD70 BSCCO at the nodal $k_F$. Figs (a) and (b) were computed using $V(q)\propto 1$ whereas for (c) and (d) Eq. \ref{eq:Vq} was used. Different curves correspond to using the entire M-EELS data (blue), keeping only the phonon peaks (orange), or keeping only the electronic continuum (green).}
	\label{fig4}
\end{figure}

For BSCCO UD70, we now analyse the role the two distinct parts of the susceptibility (Fig.~\ref{fig:fig2}b) play in the self energy. In the orange curves, the calculation uses only the phonon peaks, and in the green curves, the calculation uses only the electronic continuum. The real part of the self energy exhibits a distinct non-monotonic behaviour. We find that the 40meV and 70meV phonon peaks in the susceptibility correspond to a broad maximum and minimum at $\sim \pm60meV$ in the real part of the self-energy, and a corresponding change in slope in the imaginary part. This behaviour will be linked to the kink feature in the energy dispersion curves. When only the electronic continuum is included, there is no particular frequency scale visible in the range $[-0.2eV,0.2eV]$. As is evident from Fig.~\ref{fig4}, the qualitative aspects of these trends appear independent of the details of the interaction potential. However, as we will see in studying the spectral function, the relative ratio of the contributions of the phonons and electronic continuum is affected by the choice of the interaction potential, which has significant consequences.

\begin{figure}
	\centering
	\includegraphics[scale=0.85]{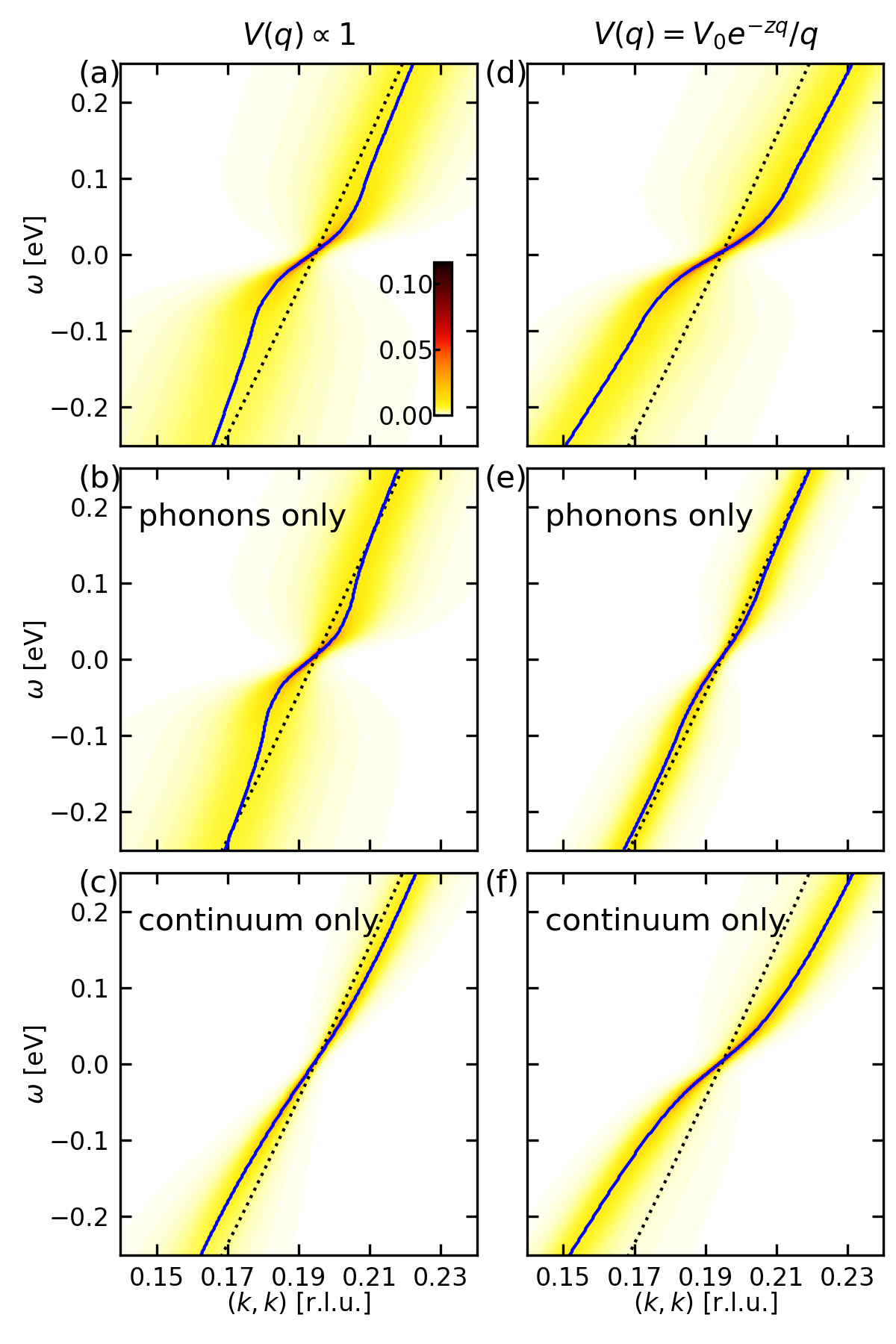}
	\caption{Spectral functions as functions of momentum and energy. Solid lines are maxima of momentum distribution curves (MDCs). Dashed lines show the non-interacting dispersion. (a)-(c) were computed for a constant potential with the full self energy (a), just the phonons (b) and just the electronic continuum (c).  Figs. (d)-(f) are the same but for the potential as shown.}
	\label{fig5}
\end{figure}

We are now set to address our key problem of the necessary ingredients to obtain the kink feature reported in previous ARPES studies on the cuprates\cite{Lanzara2001,Anzai2017,Sreedhar2020}. Fig.~\ref{fig5} contains plots of the spectral function $A(\vec{k},\omega)$ as 2d plots of frequency and momentum. To visualize the dispersion, we also plot the non-interacting dispersion (dashed lines) and the locus of maximal intensity along cuts of constant energy i.e. MDC maxima (solid lines). Panels (a)-(c) in Fig.~\ref{fig5} show the spectral functions and MDC maxima from calculations using a momentum-independent potential. As is evident, the kink feature present in panel (a) is lost completely when only the electronic continuum feature is retained. When phonons are present, there are clearly kinks at $\sim \pm60meV$ as panels Fig.~\ref{fig5}a and Fig.~\ref{fig5}b demonstrate. This is clearly correlated with the minimum and maximum in the real part of the self energy as discussed previously in Fig.~\ref{fig4}.  As seen experimentally~\cite{Lanzara2001,Anzai2017,Sreedhar2020}, the kink feature involves a renormalization of the band for $\abs{\omega} \sim 60meV$ that sharply connects back to the unrenormalized non-interacting band at larger energy. Such behaviour is absent for the strongly momentum-dependent potential. While there is a renormalization of the dispersion at low frequency in Fig.~\ref{fig5}d, there is no connection back to the non-interacting band. By analyzing the spectra when only phonons are kept (Fig.~\ref{fig5}e), we see that the phonons affect the dispersion minimally for the given form of the momentum-dependent interaction. Therefore, the renormalization in Fig.~\ref{fig5}d derives almost entirely from the contribution of the electronic continuum (Fig.~\ref{fig5}f), which has no clearly defined energy scale. While there is an apparent change in the renormalization in Fig.~\ref{fig5}d around roughly $30meV$, this is not related to any feature in $\chi(\vec{q},\omega)$ but rather controlled by the magnitude of $V_0$. (For the momentum-dependent interaction, changing this prefactor strongly affects the range where the dispersion is renormalized, in contrast to the momentum-independent interaction where the energies of the features are set by the energies of the phonon peaks.)

\begin{figure}
	\centering
	\includegraphics{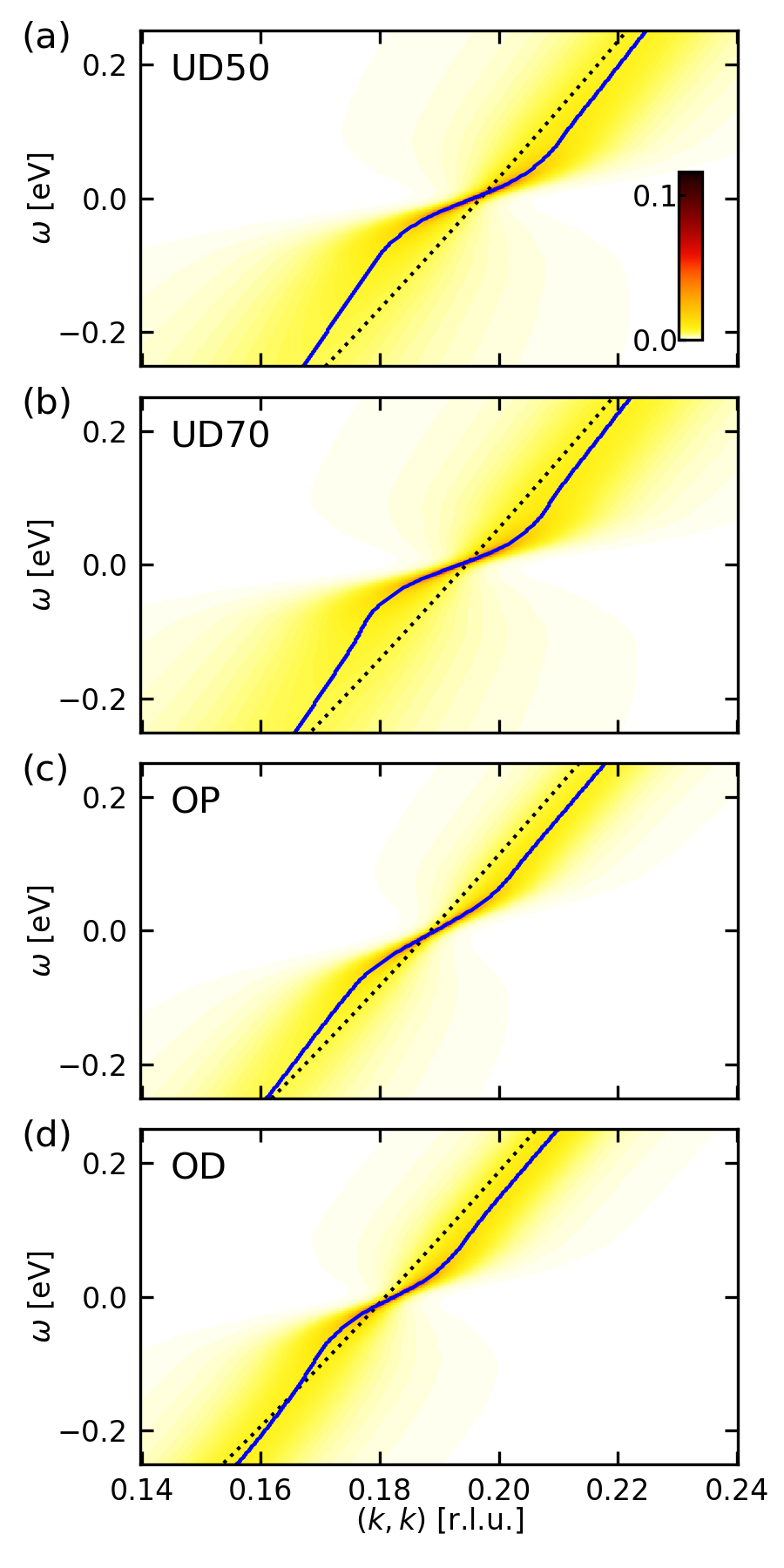}
	\caption{Spectral functions (color) and MDC maxima (solid line) calculated using the constant potential, $V({\bf q})\propto 1$ for different doping. The doping levels are as labeled starting as UD50 in (a), UD70 in (b), optimal doping (OP) in (c), and overdoping (OD) in (d).}
	\label{fig6}
\end{figure}

The doping dependence of the spectral function is shown in Fig.~\ref{fig6}. In these calculations, we have used a momentum-independent interaction and the full density response, including both the phonons and the electronic continuum. The same kink behaviour discussed previously persists in the dispersion for all doping levels though is strongest in the underdoped samples.

The crucial effect of doping on the imaginary part of the density response is the supression or enhancement of spectral weight at energies $\lesssim 0.5eV$. Relative to optimal doping, the intensity is suppressed around $\sim0.3 eV$ for overdoped samples and enhanced in under-doped samples (Fig.~\ref{fig:fig2}a). This low-energy enhancement of the susceptibility does not significantly affect the calculated spectra and the kink around 60 meV remains essentially intact.

\begin{figure}
	\centering
	\includegraphics{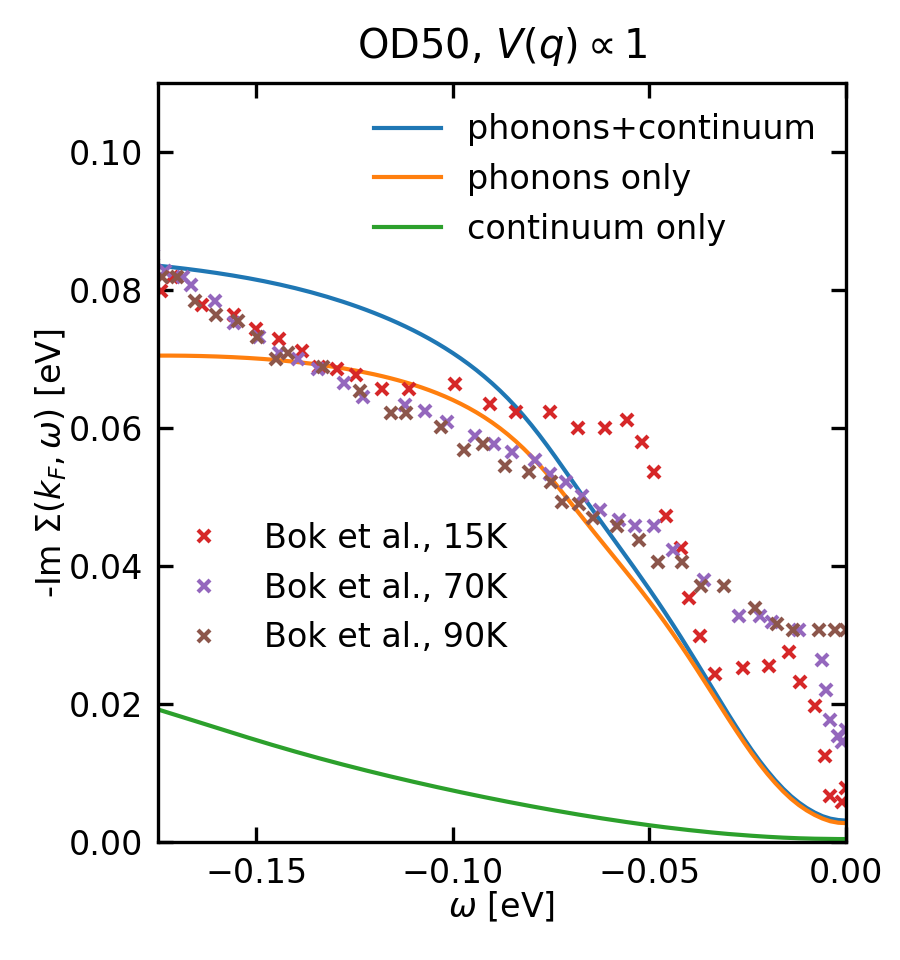}
	\caption{Comparison plots between the imaginary part of the self-energy at nodal $k_F$ obtained in this work and from ARPES for overdoped BSCCO\cite{Bok2016}. $T_c$ of the overdoped sample from Bok et al. is 82 K.}\label{fig7}
\end{figure}

As comparisons of the self-energy obtained from optical probes such as ARPES, FTIR, and ellipsometry have been made previously\cite{Puchkov1996,Kaminski2000,Damascelli2003}, it is imperative that we weigh our results in on this comparison.  Fig~\ref{fig7} plots a comparison between the imaginary parts of the self-energies obtained from our M-EELS data and ARPES~\cite{Bok2016}.  The imaginary part of the self-energy from MEELS with the constant potential is in close agreement with that extracted from ARPES\cite{Bok2016}, indicating that we have chose an appropriate strength for the interaction potential.  Once again, we show that modeling with just the electronic continuum fails to account for the sharp rise from $\omega=0$ of the self-energy.  Such behaviour can be accounted for by including just the phonon part although this contribution is peaked at $\omega=0$.  Hence, our work on extracting the self energy from M-EELS concurs with the ARPES work that phonons are the origin~\cite{Lanzara2001,Johnston2010,zaanenprl} of the kink feature at 60 meV.   While we cannot rule out other mechanisms, our work strongly suggests that phonons are sufficient to achieve a low-energy kink-like feature.  Moreover, since these features occur above $T_c$, they appear not to be dependent on superconductivity, consistent with existing literature~\cite{Damascelli2003}.








\section{Final Remarks}

Since M-EELS is inherently a 2-particle probe, it provides direct information regarding the form of the electron-electron interaction. We have found that consilience with the kink feature in ARPES requires an interaction broad in momentum space, implying the presence and importance of short-range and local repulsive interactions. Because the analysis leading to (\ref{eq:Vq}) does not strongly constrain the form of the interaction at large momenta, our result is not necessarily in contradiction with the finding that the polarizability factors in momentum and frequency, as demonstrated in (\ref{eq:Pifactor}).  

Local and short-ranged interactions are at the heart of the theoretical challenges associated with studying cuprates. Strong repulsion\cite{RMPMottness,HK1992} is the origin of Mott insulating behavior in the parent compounds and also certainly responsible for many, if not most, of the complex behavior in the doped compounds. We have shown here that these interactions are also crucial to the behavior of kinks in the single-particle dispersion, even though their existence and energy scale are tied to bosonic modes, which we identify as phonons in our M-EELS data.

Kinks due to electron-phonon coupling are typically considered in models\cite{Johnston2010} with only electron-phonon interactions. Such models display kinks in their spectra if the interactions are short-ranged or local as in, for instance, the Holstein model. (This may be contrasted, for instance, with electron-phonon interactions with forward scattering concentrated near ${\bf q} = 0$, which instead lead to replica bands\cite{Lee2014}). In the framework of the calculations in our work, these models may be understood by integrating out the phonons, leading to an effective retarded electron-electron interaction that is short-ranged and attractive. This retarded interaction gives a frequency dependence to the first term of (\ref{eq:W}) and since it contains poles at the phonon frequencies, it also yields a contribution to the self-energy that forms dispersion kinks provided the interaction is sufficiently local as we have discussed.

The Mott insulating nature of parent compound cuprates implies that the Coulombic (repulsive) electron-electron interactions significantly overcome the effective attractive interactions due to electron-phonon coupling. However, the frequency dependence of the total effective interaction remains unchanged by including repulsion, and so the first term of (\ref{eq:W}) contributes to the self-energy the same way as in a model with only electron-phonon coupling. In our simple calculations, we do not model the frequency dependence of $V({\bf q})$. Yet we find that the self-energy due to the second term of (\ref{eq:W}) alone, as in our calculations, can also lead to a substantial kink feature in $A(k,\omega)$. In this second term, the instantaneous repulsive electron-electron interactions in $V({\bf q})$ leads to kinks because of the frequency dependence of $\chi(q,\omega)$. Here, significant electron-phonon coupling is still necessary for the susceptibility to display phonon peaks as seen in the M-EELS data. The importance of all of these factors leads us to conclude that \emph{the strength of the kink feature in cuprates reflects the combined effects of strongly repulsive electron-electron interactions and electron-phonon coupling}. Such a synergistic effect has been discussed, in a different context, as a feedback loop that potentially enhances superconductivity\cite{HeYu}.

The strengthening effect of repulsive electron-electron interactions on the dispersion kink may provide a possible resolution of ab-initio calculations that have found electron-phonon coupling in cuprates to be too weak, on its own, to account for the kink observed in ARPES\cite{Giustino2007}. This effect may also be tested in simulations of microscopic lattice models involving both electron-electron and electron-phonon interactions such as the Hubbard-Holstein model. Semi-analytical diagrammatic methods similar to the GW approximation used here and Migdal-Eliashberg calculations for electron-phonon models are a possible approach. We have seen that these methods are well capable of characterizing the renormalization of the dispersion that leads to the kink feature. However we caution that such calculations do not do full justice to the strongly repulsive interactions in the cuprates. This may be seen by considering the self-energy for Mott insulators and doped Mott insulators. The simple Hatsugai-Kohmoto model\cite{HK1992} provides a clear example. In this model, a local interaction in momentum space yields the Mott physics. The single-particle dispersion, which can be solved for exactly, is modified from the non-interacting $\epsilon_k$ to $\epsilon_k+U/2 -\Sigma(k,\omega)$,
to the form
\beq
\Sigma(k,\omega)=\frac{(U/2)^2}{\omega + i0^+ - (\epsilon_k-\mu+U/2)},
\eeq
at half-filling\cite{hksupercon}. As is evident, the self-energy contains a pole at $\omega=0$, the key feature of Mottness\cite{RMPMottness} that leads to the formation of upper and lower Hubbard bands. This divergence of the self-energy cannot be captured via perturbative techniques. Therefore, we believe that non-perturbative many-body techniques and models that treat electron-electron and electron-phonon interactions on equal footing\cite{Johnston2013,Mendl2017}, while challenging, are ultimately necessary to capture the full richness of cuprate phenomenology and, in particular, the detailed behavior of dispersion kinks.

%
%
%
%
%
%
%

\textit{Acknowledgments:} EWH was supported by the Gordon and Betty Moore Foundation EPiQS Initiative through the grants GBMF 4305 and GBMF 8691. PA and PWP acknowledge support from Quantum Sensing and Quantum Materials an Energy Frontier Research Center funded by the U. S. Department of Energy, Office of Science, Basic Energy Sciences under Award DE-SC0021238. PWP also thanks the NSF DMR-1919143 for partial funding of this project. CS is supported by the DOE grant number DE-FG02-05ER46236.
\bibliographystyle{apsrev4-1}
%

\end{document}